\documentclass[aps,prd,10pt,showpacs,amsmath,showkeys,twocolumn,floatfix,amssymb, preprintnumbers, nofootinbib, superscriptaddress]{revtex4}
\usepackage{graphicx}
\usepackage{comment}
\usepackage[usenames]{color}
\usepackage{bm}
\usepackage{ifpdf}
\usepackage{floatrow}
\usepackage{makecell}
\usepackage[normalem]{ulem}
\usepackage[dvipsnames]{xcolor}
\usepackage[utf8]{inputenc}
\usepackage{hyperref}
\hypersetup{
  pdfnewwindow=true,      % links in new window
  colorlinks=true,        % false: boxed links; true: colored links
  linkcolor=PineGreen,    % color of internal links
  citecolor=PineGreen,    % color of links to bibliography
  filecolor=PineGreen,    % color of file links
  urlcolor=PineGreen      % color of external links
}
\newcommand{\be}{\begin{eqnarray}}
\newcommand{\ee}{\end{eqnarray}}

\begin{document}

\title{Coulomb corrections to two-particle interaction in artificial traps}

\author{Peng~Guo}
\email{pguo@csub.edu}

\affiliation{Department of Physics and Engineering,  California State University, Bakersfield, CA 93311, USA}
\affiliation{Kavli Institute for Theoretical Physics, University of California, Santa Barbara, CA 93106, USA}

\date{\today}

\begin{abstract}
In present work, we discuss the effect of Coulomb interaction to the dynamics of two-particle system bound in various traps. The  strategy of including Coulomb interaction into the quantization condition of trapped system is discussed in a general and non-perturbative manner.  In most cases,  Coulomb corrections to quantization condition     largely rely on numerical approach or perturbation expansion. Only for some special cases, such as the spherical hard wall trap, a closed-form of quantization condition with all orders of Coulomb corrections can be obtained.
\end{abstract}

\maketitle

\section{Introduction}\label{sec:intro}

  Recent advances in lattice quantum Chromodynamics (LQCD),  {\it ab initio}  nuclear many-body theory and developments in computer technology have now made it possible for the high precision  computation of hadron and nuclei systems from the first principle. However, most of these computations are performed in various traps, for instance, harmonic oscillator trap in nuclear physics and periodic cubic box in LQCD. The typical observables from these {\it ab initio}   computations are  discrete energy spectrum of trapped systems. Therefore,  extracting particle interactions from discrete energy spectrum in the trap and building connection between trapped dynamics and infinite volume dynamics have became an important subject in both LQCD and nuclear physics communities in recent years. In elastic two-particle sector, such a connection between trapped system and infinite volume system can be formulated in a closed form, such as  L\"uscher formula  \cite{Luscher:1990ux} in a periodic cubic box in LCQD  and BERW formula \cite{Busch98} in a harmonic oscillator trap in nuclear physics community. Since then, L\"uscher and BERW  formula have been quickly extended into both coupled-channel and  few-body sectors, see e.g. Refs.~\cite{Rummukainen:1995vs,Christ:2005gi,Bernard:2008ax,He:2005ey,Lage:2009zv,Doring:2011vk,Guo:2012hv,Guo:2013vsa,Kreuzer:2008bi,Polejaeva:2012ut,Hansen:2014eka,Mai:2017bge,Mai:2018djl,Doring:2018xxx,Guo:2016fgl,Guo:2017ism,Guo:2017crd,Guo:2018xbv,Mai:2019fba,Guo:2018ibd,Guo:2019hih,Guo:2019ogp,Guo:2020wbl,Guo:2020kph,Guo:2020iep,Guo:2020ikh,Guo:2020spn,Guo:2021lhz,Guo:2021uig}. Both L\"uscher and BERW  formula have the form of 
  \begin{equation}
   \det \left [  \cot \delta  (E) -  \mathcal{M} (E ) \right ]=0 \, ,
    \label{eqn:generalQC}
\end{equation}
where  $\delta (E)$ refers to the diagonal matrix of scattering phase shifts, and the analytic matrix function
$ \mathcal{M} (E ) $  is associated to the geometry and dynamics of trap itself. L\"uscher and BERW  formula  as the matter of fact is the result of    the presence of two well separated physical  scales: (1) short-range interaction between two particles and (2) size of trap. Hence the short-range dynamics that is described by scattering phase shift  and long-range correlation effect due to the  trap can be factorized.

The aim of present work is to extend such a relation to include long-range Coulomb interaction between charged particles.   Coulomb interaction  becomes dominant for charged particles interactions at low energy \cite{Kong:1999sf},  including Coulomb interaction may be crucial for charged system interaction in a trap,  see e.g. charge hadron system in LQCD \cite{Beane:2020ycc}.  In fact,  some early works on including Coulomb corrections in finite volume has already been presented in Refs.~\cite{Beane:2014qha,Stellin:2020gst}. The discussion   in Refs.~\cite{Beane:2014qha,Stellin:2020gst} was primarily  based on effective perturbation field theory approach. It has been well known fact  that  both incoming plane wave and scattered spherical wave are distorted by  long-range Coulomb interaction \cite{messiah1999quantum},  
\begin{align}
    \psi^{(\infty)}_{l }(r,q)    
&  \stackrel{r\rightarrow \infty}{\sim }   \frac{     \sin ( q r - \frac{\pi}{2} l +    \frac{Z \mu}{q} \ln 2 q r ) }{ q r}  \nonumber \\
& +    t_l  (q)    \frac{e^{ i (q r - \frac{\pi}{2} l  +  \frac{Z \mu}{q}  \ln 2 q r) }}{q r}      ,
\end{align}
where $Z= - Z_1 Z_2 e^2$ is Coulomb interaction strength, and $\mu$ and $q$ refers to the effective mass and incoming momentum of two-particle system. $t_l (q)$ is the partial wave scattering amplitude. Hence perturbation breaks down and Coulomb corrections must be dealt with  non-perturbatively, see e.g.  \cite{Kong:1999sf}. When it comes to formulating   L\"uscher and BERW  formula   in the presence of long-range Coulomb force,  Coulomb propagator must be used instead of free particle propagator. In this work, we offer a general  perspective for the formulating L\"uscher and BERW  formula   in   presence of long-range Coulomb force. All the discussion are based on Lippmann-Schwinger (LS) equation approach, hence, the discussion can be made general in non-perturbative way for various types of trap. However, except the hard-sphere wall trap, the analytic form of Green's function in a trap is usually not available, Dyson equation must be solved either numerically or in terms of perturbation expansion.

The paper is organized as follows.  The derivation of the Coulomb corrections to the quantization condition of trapped system is  presented in Sec.~\ref{trapscatt}. The discussions and summary are given in Sec.~\ref{summary}.

\section{Connecting bound states in a trap to infinite volume scattering state with Coulomb force}\label{trapscatt}

 In this section, we present a   general formalism  on the topic of bridging discrete  bound state energy spectrum in a trap and infinite volume scattering  dynamics in the presence of Coulomb force. The     commonly used   traps  are periodic finite box in LQCD \cite{Beane:2020ycc}, harmonic potential in nuclear physics \cite{Rotureau:2010uz, Rotureau:2011vf, Luu:2010hw, Zhang:2020rhz}, and spherical hard wall in some of the lattice implementations of chiral effective field theory  \cite{Elhatisari:2016hby,Rokash:2015hra}. The    brief discussion of  formal scattering in presence of both a short-range interaction and a long-range Coulomb interaction is   given in Appendix~\ref{scattcoulinf}.  Before the technical  presentation of  detailed derivation of quantization conditions, our notations for describing the dynamics of two-particle interaction  in trap and in infinite volume   are established as follows:

\paragraph{Dynamics in a trap: }  the relative motion of  two charged spinless  particles interacting with both Coulomb and short-range interactions in a trap is described by Schr\"odinger equation
 \begin{equation}
\left [\varepsilon -\hat{H}_{t} - V_C (r) \right ] \psi^{(t)}_{\varepsilon}(\mathbf{ r}) = \int_{trap} d \mathbf{ r}'  V_S (\mathbf{ r} , \mathbf{ r}')   \psi^{(t)}_{\varepsilon}(\mathbf{ r}')   , \label{schtrap}
 \end{equation}
 where $\int_{trap} d \mathbf{ r}'$ refers to the integral over  space of unit cell of the trap, {\it e.g.}
 $$\int_{trap} d \mathbf{ r}' = \int^{\frac{L}{2}}_{- \frac{L}{2}} d x' d y' d z'$$ in a periodic box with the size of $L$.
  The Hamiltonian operator of the trap is given by  
 \begin{equation}
 \hat{H}_{t} = \hat{H}_{0} +V_{trap} (\mathbf{ r})  ,
 \end{equation}
 with
     \begin{equation}
      \hat{H}_{0}= - \frac{\nabla_\mathbf{ r}^2}{2\mu} 
       \end{equation}
      and $V_{trap} (\mathbf{ r}) $ representing the free particle Hamiltonian operator and trap potential respectively. $\mu$ stands for the reduced mass of two particles, and $\varepsilon$ is  energy of trapped  particles  associated with relative motion.
 \begin{equation}
 V_C (r) = - \frac{Z}{r}
 \end{equation}
  and  $V_S (\mathbf{ r}, \mathbf{ r}' )$ denote the Coulomb and short-range interactions between particles respectively.

\paragraph{Dynamics in infinite volume: }     the dynamics of two charged interacting particles through the same short-range interaction $V_S (\mathbf{ r}, \mathbf{ r}' )$ in infinite volume   is given by
  \begin{equation}
 \left [\varepsilon_\infty -  \hat{H}_{0} - V_C (r)  \right ] \psi^{( \infty)}_{\varepsilon_\infty }(\mathbf{ r}) =  \int_{-\infty}^\infty d \mathbf{ r}'  V_S (\mathbf{ r} , \mathbf{ r}')    \psi^{( \infty)}_{\varepsilon_\infty}(\mathbf{ r}')   , \label{schinf}
 \end{equation}
where $\varepsilon_\infty $ stands for the relative motion  energy of particles in infinite volume.    $\varepsilon_\infty $ is related to   $\varepsilon$ in the trap by total energy conservation,
 \begin{equation}
 \varepsilon_\infty + \frac{\mathbf{ P}^2}{2M} = \varepsilon + E^{(t)}_{ CM} =E,
 \end{equation} 
 where  $\frac{\mathbf{ P}^2}{2M} $  and $E^{(t)}_{ CM} $  are the center of mass (CM) energy of system in  infinite volume   and in the trap respectively.

\paragraph{Short-range interaction: }       the separable zero-range potential is assumed in follows for the  derivation of quantization condition.  In coordinate space, it has the form  of, see Refs.\cite{Guo:2021lhz,Guo:2021uig}, 
 \begin{equation}
 V_S(\mathbf{ r}, \mathbf{ r}' )  = \frac{ \delta(r)  \delta(r')  }{ (rr')^2}  \sum_{l m}    \frac{V^{(S)}_l }{   (r r')^l } Y_{lm} (\mathbf{ \hat{r}} )Y^*_{lm} (\mathbf{ \hat{r}}' ). \label{Vcord}
 \end{equation}
 We emphasis that the assumption of separable zero-range potential is not essential for obtaining the L\"uscher  or BERW formula  type quantizations condition, due to the fact that  L\"uscher  or BERW formula  type quantization conditions are model independent asymptotic result when the size of trap is much larger than the range of nuclear interactions, see {\it e.g.}   discussion in Refs.\cite{Guo:2021lhz,Guo:2021uig}. However, separable zero-range potential does serve as a convenient tool for the derivation of quantization condition.

Next, the dynamical equations of charged particles interaction in a trap  is presented  in \ref{dynamicstrap},   and   quantization condition of trapped particles system that connects the  strength of   short-range interaction and the Coulomb Green's function in a trap  is also derived and given  in \ref{dynamicstrap}.  Then, under the same assumption of separable short-range interaction, the scattering solutions   of charged particles in infinite volume are given in details in \ref{dynamicsinf}, the similar relation that connecting the strength of short-range interaction, infinite volume Coulomb Green's function and scattering phase shift is also obtained. At last, by combining dynamical equations in a trap and infinite volume together, the L\"uscher  or BERW formula  type of quantization condition   is obtained and given  in \ref{dynamicsQC}.

\subsection{Coulomb force modified  dynamical equations in a trap}\label{dynamicstrap} 
 In the trap, the integral representation of Eq.(\ref{schtrap}) is given by  
\begin{align}
 \psi^{(t)}_{\varepsilon}(\mathbf{ r})  &= \int_{trap} d \mathbf{ r}'' G^{(C,t)} (\mathbf{ r}, \mathbf{ r}'' ; \varepsilon ) \nonumber \\
 & \times   \int_{trap} d \mathbf{ r}' V_S (\mathbf{ r}'',\mathbf{ r}')   \psi^{(t)}_{\varepsilon}(\mathbf{ r}')   , 
\end{align}  
where 
\begin{equation}
G^{(C,t)} (\mathbf{ r}, \mathbf{ r}'' ; \varepsilon) = \langle\mathbf{ r}  | \frac{1}{ \varepsilon - \hat{H}_{t} - \hat{V}_C  }  | \mathbf{ r}'' \rangle  
\end{equation}
stands for the Coulomb Green's function in a trap. The Coulomb Green's function $G^{(C,t)} $ satisfies Dyson equation,
\begin{align}
&G^{(C,t)} (\mathbf{ r}, \mathbf{ r}'' ; \varepsilon) = G^{(t)} (\mathbf{ r}, \mathbf{ r}'' ; \varepsilon)  \nonumber \\
&+   \int_{trap} d \mathbf{ r}'  G^{(t)} (\mathbf{ r}, \mathbf{ r}' ; \varepsilon)  V_C( r') G^{(C,t)} (\mathbf{ r}', \mathbf{ r}'' ; \varepsilon) , \label{DysonTCGF}
\end{align}
where
\begin{equation}
G^{(t)} (\mathbf{ r}, \mathbf{ r}'' ; \varepsilon) = \langle\mathbf{ r}  | \frac{1}{ \varepsilon - \hat{H}_{t}  }  | \mathbf{ r}'' \rangle  
\end{equation}
is     particle Green's function in a trap. The partial wave expansions
\begin{equation}
  \psi^{(t)}_{\varepsilon}(\mathbf{ r})  = \sum_{ l m }  \psi^{(t)}_{lm }(r)  Y_{lm} (\mathbf{ \hat{r}}) 
  \end{equation}
  and
  \begin{align}
& G^{(C,t)} (\mathbf{ r}, \mathbf{ r}'' ; \varepsilon)  \nonumber \\
& = \sum_{lm, l''m''}  Y_{lm} (\mathbf{ \hat{r}}) G_{ lm, l''m''}^{(C,t)} (r, r'' ; \varepsilon) Y^*_{l''m''} (\mathbf{ \hat{r}}'')
\end{align}
 yields
\begin{align}
 \psi^{(t)}_{lm }(r)  &= \sum_{l' m'} \int_{trap}  {r''}^2 d r'' G_{ lm, l'm'}^{(C,t)} (r, r'' ; \varepsilon) \nonumber \\
 & \times   \int_{trap} {r'}^2 d r' V^{(S)}_{l'} (r'', r')   \psi^{(t)}_{l'm'}(r')   . \label{LStrapPWA}
\end{align}  
With separable potential given in Eq.(\ref{Vcord}),   the quantization condition that determines the discrete bound state energy spectrum of trapped system is thus  given by
 \begin{equation}
  \det   \left [ \delta_{lm, l'm'}  \frac{1 }{  V^{(S)}_l} -     \frac{G_{ lm, l'm'}^{(C,t)} (r, r' ; \varepsilon) }{   r^l  {r' }^{l'} } |_{r,r'\rightarrow 0}  \right ]   =0 . \label{QCVtrap}
\end{equation}  

The  Coulomb Green's function, $G^{(C,t)}$, that describes the propagation of  charged particles in a trap is an essential ingredient of quantization condition, and   must be solved   first. Specifically, only commonly used traps in lattice and nuclear physics communities are considered in this work:

\subsubsection{Harmonic oscillator trap}
In the harmonic oscillator (HO) trap with trap potential: 
\begin{equation}
V_{trap} (r) = \frac{1}{2} \mu \omega^2 r^2 ,
\end{equation}
 the rotational symmetry is still preserved, thus only diagonal elements of partial wave Green's function contribute. The partial wave Dyson equation for    harmonic trap Green's function in presence of Coulomb force is given by
\begin{align}
&G_l^{(C,\omega)} (r, r' ; \varepsilon) = G^{(\omega)}_l (r, r' ; \varepsilon)  \nonumber \\
&-   \int_{0}^\infty {r''}^2 d  r''  G_l^{(\omega)} (r, r'' ; \varepsilon)  \frac{Z}{r''}  G_l^{(C,\omega)} (r'', r' ; \varepsilon) , \label{DysonTCGFHO}
\end{align}
where   $G^{(  \omega)}_l$ is the partial-wave HO Green's function and is given in Refs.~\cite{Blinder83,Guo:2021uig} by
 \begin{align}
 & G^{(  \omega)}_l (r, r'; \varepsilon )  = - \frac{1}{\omega (r r')^{ \frac{3}{2}}} \frac{ \Gamma (\frac{l}{2} + \frac{3}{4} - \frac{ \varepsilon }{2 \omega}) }{\Gamma(l+\frac{3}{2})}  \nonumber \\
& \times  \mathcal{M}_{\frac{\varepsilon }{2\omega}, \frac{l}{2} + \frac{1}{4} }( \mu \omega r^2_{<}) \mathcal{W}_{\frac{\varepsilon }{2 \omega},\frac{l}{2} + \frac{1}{4} }  (\mu \omega r^2_{>}) .
 \end{align}
$\mathcal{ M}_{a,b}(z)$ and $\mathcal{W}_{a,b}(z)$ are the Whittaker functions as defined in  Ref.~\cite{NIST:DLMF}, and $r_<$ and $r_>$ represent the lesser and  greater of $(r,r')$ respectively.

\subsubsection{Periodic  cubic box} 
In finite volume, the trap potential is replaced by periodic boundary condition. The rotational symmetry is broken, and angular orbital momenta are no longer good quantum numbers. In addition,  the periodic boundary condition is not satisfied by infinite volume Coulomb potential: $V_C(r) = - \frac{Z}{r}$.   The infinite volume Coulomb potential is usually replaced by infrared singularity regularized periodic Coulomb potential, see Refs.~\cite{Beane:2014qha,Stellin:2020gst},
\begin{equation}
V_{C}^{(L)} (\mathbf{ r}) = -  \frac{1}{L^3} \sum_{\mathbf{ p} = \frac{2\pi \mathbf{ n}}{L} , \mathbf{ n} \in \mathbb{Z}^3, \mathbf{ n} \neq \mathbf{ 0} } \frac{4\pi Z}{|\mathbf{ p} |^2} e^{i\mathbf{  p} \cdot \mathbf{ r}},
\end{equation}
where $L$ is size of cubic box, and 
\begin{equation}
V_{C}^{(L)} (\mathbf{ r}+ \mathbf{ n} L) = V_{C}^{(L)} (\mathbf{ r}), \ \ \ \ \mathbf{ n} \in \mathbb{Z}^3 .
\end{equation}
In momentum space, Dyson equation in finite volume is given by
\begin{align}
&\widetilde{G}^{(C,L)} (\mathbf{ p}, \mathbf{ p}' ; \varepsilon)  = \frac{L^3 \delta_{\mathbf{ p} , \mathbf{ p}'}}{\varepsilon - \frac{\mathbf{ p}^2}{2\mu}} \nonumber \\
&-  \frac{1}{\varepsilon - \frac{\mathbf{ p}^2}{2\mu}}  \frac{1}{L^3} \sum_{\mathbf{ p}'' = \frac{2\pi \mathbf{ n}}{L} , \mathbf{ n} \in \mathbb{Z}^3 }^{ \mathbf{ p}'' \neq \mathbf{ p}} \frac{4\pi Z}{|\mathbf{ p} - \mathbf{ p}''|^2} \widetilde{G}^{(C,L)} (\mathbf{ p}'', \mathbf{ p}' ; \varepsilon). \label{DysonTCGFFV}
\end{align}
 The finite volume Coulomb force modified Green's function in coordinate space is thus given by finite volume Fourier transform
 \begin{equation}
G^{(C,L)} (\mathbf{ r}, \mathbf{ r}' ; \varepsilon) = \frac{1}{L^6} \sum_{\mathbf{ p}, \mathbf{ p}' \in \frac{2\pi \mathbf{ n}}{L}  }^{\mathbf{ n} \in \mathbb{Z}^3} e^{ i \mathbf{ p} \cdot \mathbf{ r} }\widetilde{G}^{(C,L)} (\mathbf{ p}, \mathbf{ p}' ; \varepsilon) e^{ -i \mathbf{ p}' \cdot \mathbf{ r}' }.
 \end{equation}

\subsubsection{Spherical hard wall} 
The hard-sphere boundary condition is accomplished by the trap potential
\begin{equation}
V_{trap} (r)= \begin{cases} 0, & r<R  \, , \\ \infty, & r>R\, ,  \end{cases}
\end{equation}
where $R$ is the radius of the sphere. Hence, inside of spherical hard wall: $|\mathbf{ r}| <R$, Coulomb force modified Green's function satisfies
\begin{equation}
\left [ \varepsilon - \hat{H}_{0} - \hat{V}_C  \right ]G^{(C,h.s.)} (\mathbf{ r}, \mathbf{ r}'; \varepsilon) = \delta(  \mathbf{ r}  - \mathbf{ r}' )  ,
\end{equation}
which is just regular   differential equation for Coulomb Green's function except boundary condition.

\subsection{Coulomb force modified  infinite volume dynamical equations}\label{dynamicsinf}  
In infinite volume,   
 the scattering solution of two charged interacting particles   in presence of Coulomb interaction is described by inhomogeneous   LS equation,  
\begin{align}
 & \psi^{(\infty)}_{ \varepsilon_\infty}(\mathbf{ r},\mathbf{ q})  = \psi^{(C,\infty)}_{ \varepsilon_\infty}(\mathbf{ r},\mathbf{ q})    \nonumber \\
 & + \int_{ -\infty}^\infty d \mathbf{ r}'' G^{(C,\infty)} (\mathbf{ r}, \mathbf{ r}'' ; q)   \int_{-\infty}^\infty d \mathbf{ r}' V_S (\mathbf{ r}'',\mathbf{ r}')   \psi^{(\infty)}_{ \varepsilon_\infty}(\mathbf{ r}',\mathbf{ q})    , 
\end{align}  
where $ \mathbf{ q} $ is on-shell incoming momentum: 
\begin{equation}
q = \sqrt{2\mu \varepsilon_\infty}.
\end{equation}
  $ \psi^{(C,\infty)}_{ \varepsilon_\infty}   $ and $G^{(C,\infty)}$  are Coulomb wave function and Coulomb Green's function respectively.   The partial wave expansion  
\begin{align}
& \psi^{(\infty)}_{ \varepsilon_\infty}(\mathbf{ r},\mathbf{ q})  = \sum_{l m}  Y^*_{lm} (\mathbf{ \hat{q}}) \psi^{(\infty)}_{  l}(r,q) Y_{lm} (\mathbf{ \hat{r}}), \nonumber \\
& G^{(C,\infty)} (\mathbf{ r} , \mathbf{ r}'' ; q) = \sum_{lm }  Y_{lm} (\mathbf{ \hat{r}}) G_{ l}^{(C, \infty)} (r, r'' ; q) Y^*_{lm} (\mathbf{ \hat{r}}'') , \label{GinfPW}
\end{align}
and separable potential in Eq.(\ref{Vcord}) yield an algebra equation
\begin{align}
 & \frac{\psi^{(\infty)}_{l }(r,q) }{r^l} = \frac{\psi^{(C,\infty)}_{l }(r,q) }{r^l}  \nonumber \\
& +  V^{(S)}_l   \frac{G_{ l }^{(C,\infty)} (r, r'' ; q)  }{   (r r'')^l }    \frac{ \psi^{(\infty)}_{l}(r',q)  }{{r'}^l} |_{r',r''\rightarrow 0}   .  \label{LSinfcoulomb}
\end{align}

\subsubsection{Coulomb wave function and Coulomb Green's function} 
The analytic expression of $ \psi^{(C,\infty)}_{l}   $ and $G_l^{(C,\infty)}$  are given in Refs.~\cite{messiah1999quantum,doi:10.1063/1.1704153} respectively by
\begin{align}
& \psi^{(C,\infty)}_{  l}(r,q)  =4\pi\frac{\Gamma(l+1+i \gamma)}{(2 l+1)!} e^{- \frac{\pi}{2} \gamma}  \nonumber \\
& \times (2 i q r)^l  e^{i q r} M(l+1+ i \gamma, 2L+2, -2 i q r),
\end{align}
and
\begin{align}
& G_{ l }^{(C,\infty)} (r, r'' ; q) =2 \mu (2 i q) \frac{\Gamma(l+1 + i \gamma)}{(2l+1)!}  \nonumber \\
& \times (-2 i q r_<)^l  e^{i q r_<} M(l+1+ i \gamma, 2l+2, -2 i q r_<) \nonumber \\
& \times (-2 i q r_>)^l  e^{i q r_>} U(l+1+ i \gamma, 2l+2, -2 i q r_>),
\end{align}
where  $M(a,b,z) $ and $U(a,b,z)$ are two linearly independent Kummer functions \cite{NIST:DLMF}, and
\begin{equation}
\gamma = - \frac{ Z \mu}{q}.
\end{equation}

For the convenience, let's introduce two real functions:
\begin{align}
& j_l^{(C)}(\gamma, qr)= C_l (\gamma)     (q r)^l  e^{i q r} M(l+1+ i \gamma, 2l+2, -2 i q r), \label{jLC}
\end{align}
and
\begin{align}
& n_l^{(C)}(\gamma, q r)  \nonumber \\
&=  i(-2q r)^l e^{ \frac{\pi}{2}\gamma}     e^{i q r} U(l+1+ i \gamma, 2l+2, -2 i q r) e^{ i \delta_l^{(C)}}    \nonumber \\
 &-i(-2q r)^l e^{ \frac{\pi}{2}\gamma}      e^{-i q r} U(l+1- i \gamma, 2l+2, 2 i q r) e^{- i \delta_l^{(C)}}  , \label{nLC}
\end{align}
where the Sommerfeld factor and Coulomb phase shift are defined in \cite{messiah1999quantum} by
\begin{equation}
C_l (\gamma) = 2^l \frac{|\Gamma(l+1+ i \gamma)|}{(2 l+1)!} e^{- \frac{\pi}{2} \gamma },
\end{equation}
and
 \begin{equation}
e^{2 i \delta_l^{(C)}}  =   \frac{\Gamma(l+1+ i \gamma)}{\Gamma(l+1- i \gamma) }  . \label{phasecoulomb}
\end{equation}
At the limit of $\gamma \rightarrow 0$, $ j_l^{(C)}(\gamma, qr)$ and $ n_l^{(C)}(\gamma, qr)$   are reduced to the regular spherical Bessel functions,
\begin{equation}
\left ( j_l^{(C)}(\gamma, qr) , n_l^{(C)}(\gamma, qr) \right ) \stackrel{\gamma \rightarrow 0}{\rightarrow } \left  ( j_l( qr) , n_l( qr) \right ).
\end{equation}
Also using identity
\begin{align}
&M(l+1+ i \gamma, 2l+2, -2 i q r)   \nonumber \\
&= - (-1)^l \frac{(2 l+1)!}{\Gamma(l+1 - i \gamma)} e^{ \pi \gamma} U(l+1+ i \gamma, 2l+2, -2 i q r) \nonumber \\
& - (-1)^l \frac{(2 l+1)!}{\Gamma(l+1 + i \gamma)} e^{ \pi \gamma} e^{- 2 i q r} U(l+1- i \gamma, 2l+2, 2 i q r) ,
\end{align}
the partial wave Coulomb wave function and Coulomb Green's function can thus be rewritten as
\begin{equation}
 \psi^{(C,\infty)}_{  l}(r,q)  =4\pi  i^l j_l^{(C)}(\gamma, qr)  e^{ i \delta_l^{(C)}}  ,\label{purecoulombwav}
\end{equation}
and
\begin{equation}
G_{ l }^{(C,\infty)} (r, r'' ; q)    = - i 2 \mu  q j_l^{(C)} ( \gamma, q r_<)   h_l^{(C,+)} (\gamma, q r_>)  , \label{Gcoulombinf}
\end{equation}
where
\begin{align}
 & h_l^{(C,\pm)} (\gamma, q r) =  j_l^{(C)} (\gamma, q r)  \pm i  n_l^{(C)} (\gamma, q r) \nonumber \\
 & = -2   (-2q r)^l   e^{ \frac{\pi}{2} \gamma}  e^{ \pm i q r} U(l+1\pm i \gamma, 2l+2, \mp 2 i q r) e^{\pm i \delta_l^{(C)}}   .
\end{align}
The Coulomb Green's function in Eq.(\ref{Gcoulombinf})  thus resemble the free particle Green's function,
\begin{equation}
G_{ l }^{(0,\infty)} (r, r'' ; q)    = - i 2 \mu  q j_l  (  q r_<)   h_l^{(+)} ( q r_>)  ,
\end{equation}
where $j_l$ and $h_l^{(+)}$ are regular spherical Bessel and Hankel functions.

\subsubsection{Coulomb force modified scattering amplitudes }
In presence of Coulomb force, the total scattering amplitude now is composed of two components: (1)  the short-range interaction scattering amplitude modified by Coulomb interaction and  (2) the pure Coulomb scattering amplitude.

\paragraph{Coulomb force modified short-range interaction scattering amplitude:} the short-range interaction scattering amplitude can be defined by the solution of Eq.(\ref{LSinfcoulomb}),
\begin{align}
   \psi^{(\infty)}_{l }(r,q) &  = 4\pi i^l     \bigg [    j^{(C)}_l(\gamma, q r)e^{ i \delta_l^{(C)} }   \nonumber \\
& +     i t^{(SC)}_l(q)   h_l^{(C,+)} (\gamma, q r)e^{- i \delta_l^{(C)} }   \bigg ] ,  \label{totwavinf}
\end{align}
where  $t^{(SC)}_l(q) $ is the Coulomb force modified  short-range interaction scattering amplitude and is given by
\begin{equation}
 t^{(SC)}_l(q)   =  -  \frac{ 2 \mu q   \left ( \frac{j_l^{(C)}(\gamma, qr)  }{r^L} |_{r\rightarrow 0} \right )^2 }{  \frac{1}{ V^{(S)}_l } -   \frac{G_{ l }^{(C,\infty)} (r', r'' ; q)  }{   (r' r'')^l }|_{r',r'' \rightarrow 0}   }  e^{2 i \delta_l^{(C)}}    .   \label{TLinf}
\end{equation} 
The $t^{(SC)}_l(q) $ is typically parameterized by both Coulomb phase shift $\delta_l^{(C)}$ and a short-range scattering phase shift $\delta_l^{(S)}$,
\begin{equation}
 t^{(SC)}_l(q)    = \frac{1}{\cot \delta_l^{(S)} - i}  e^{2 i \delta_l^{(C)}}. \label{TSCparam}
\end{equation}
Using the asymptotic form:
\begin{align}
 \frac{j_l^{(C)}(\gamma, qr)  }{r^l} |_{r\rightarrow 0}  &=  q^l C_l(\gamma) , \nonumber \\
  \Im \left [ \frac{G_{ l }^{(C,\infty)} (r', r'' ; q)  }{   (r' r'')^l }|_{r',r'' \rightarrow 0}  \right ] & = - 2 \mu  q^{2 l+1}  C_l^2(\gamma)  ,
\end{align}
and Eq.(\ref{TLinf}) and Eq.(\ref{TSCparam}), 
one thus find 
\begin{align}
  \frac{1}{ V^{(S)}_l } &= - 2 \mu q^{2 l+1} C^2_l (\gamma) \cot \delta_l^{(S)} (q)  \nonumber \\
& + \Re \left [ \frac{G_{ l }^{(C,\infty)} (r', r'' ; q)  }{   (r' r'')^l }|_{r',r'' \rightarrow 0}  \right ]. \label{Vcotdeltainf}
\end{align}

\paragraph{Pure Coulomb scattering amplitude:}  the pure Coulomb scattering amplitude is defined by Coulomb wave function.  By introducing 
\begin{equation}
  h_l^{(C,\pm)} (\gamma, q r)  = e^{\pm i \delta_l^{(C)}}  H_l^{(C,\pm)} (\gamma, q r)   ,
\end{equation}
where
\begin{align}
 & H_l^{(C,\pm)} (\gamma, q r) =  J_l^{(C)} (\gamma, q r)  \pm i  N_l^{(C)} (\gamma, q r) \nonumber \\
 & = -2   (-2q r)^l   e^{ \frac{\pi}{2} \gamma}  e^{ \pm i q r} U(l+1\pm i \gamma, 2l+2, \mp 2 i q r)   ,
\end{align}
we can thus rewrite the Coulomb wave function  in Eq.(\ref{purecoulombwav}) to
\begin{equation}
 \psi^{(C,\infty)}_{  l}(r,q)   
=4\pi  i^l  \left [ J_l^{(C)} (\gamma, q r) +    i t_l^{(C)} (q)  H_l^{(C,+)} (\gamma, q r) \right ],
\end{equation}
where  $t_l^{(C)} (q) $ is the pure Coulomb scattering amplitude:
\begin{equation}
t_l^{(C)} (q) =  \frac{e^{ 2 i \delta_l^{(C)}} -1}{2i } .
\end{equation}

\paragraph{Total scattering amplitude:}   the total wave function in Eq.(\ref{totwavinf}) is now also given by
\begin{align}
   \psi^{(\infty)}_{l }(r,q) &  = 4\pi i^l     \bigg [    J^{(C)}_l(\gamma, q r)   +     i    t_l(q)   H_l^{(C,+)} (\gamma, q r)  \bigg ]  ,
\end{align}
where
\begin{align}
  t_l(q)   = t^{(C)}_l(q) +t^{(SC)}_l(q)  = \frac{e^{2 i \delta_l^{(C)}} e^{2 i \delta_l^{(S)}} -1}{2 i}.
\end{align}

\paragraph{Asymptotic forms of wave functions:}    using asymptotic form of $ H_l^{(C,\pm)} $ functions,
\begin{equation}
   H_l^{(C,\pm)} (\gamma, q r)    
     \stackrel{ r\rightarrow \infty}{ \rightarrow}   h^{(\pm)}_l ( qr)   e^{ \mp i    \gamma \ln 2 q r } ,
\end{equation}
one can easily illustrate that
\begin{align}
 \psi^{(C,\infty)}_{  l}(r,q)  
&   \stackrel{ r\rightarrow \infty}{ \rightarrow}    4\pi  i^l  \bigg [  \frac{     \sin ( q r - \frac{\pi}{2} l -    \gamma \ln 2 q r ) }{ q r}  \nonumber \\
& + i t_l^{(C)} (q)    h^{(+)}_l ( qr)   e^{ - i    \gamma \ln 2 q r } \bigg ]    ,
\end{align}
and  
\begin{align}
    \psi^{(\infty)}_{l }(r,q)    
&   \stackrel{ r\rightarrow \infty}{ \rightarrow}    4\pi  i^l  \bigg [  \frac{     \sin ( q r - \frac{\pi}{2} l -    \gamma \ln 2 q r ) }{ q r}  \nonumber \\
& + i   t_l  (q)   h^{(+)}_l ( qr)   e^{ - i    \gamma \ln 2 q r } \bigg ]    ,
\end{align}
where the factor $ \gamma \ln 2 q r$ represents the long-range Coulomb distortion  effect to the incoming  plane wave and outgoing spherical wave.

\subsection{Quantization condition in a trap in presence of Coulomb interaction}\label{dynamicsQC}
Combining Eq.(\ref{QCVtrap}) and Eq.(\ref{Vcotdeltainf}) by eliminating $V^{(S)}_l$, one thus find
 Coulomb force modified  L\"uscher formula-like relation,
 \begin{equation}
  \det   \left  [  \delta_{lm, l'm'}        \cot \delta^{(S)}_l (q)  - \mathcal{M}^{(C,t)}_{lm, l'm'}  (\varepsilon)    \right  ]   =0 ,
\end{equation}  
where $\mathcal{M}^{(C,t)}$ is generalized zeta function in presence of Coulomb force,
 \begin{align}
  &  \mathcal{M}^{(C,t)}_{lm, l'm'}  (\varepsilon)   = -\frac{1}{ 2\mu q^{2l+1}  C^2_l (\gamma) }   \frac{G_{ lm, l'm'}^{(C,t)} (r, r' ; \varepsilon) }{   r^l  {r' }^{l'} } |_{r,r'\rightarrow 0}  \nonumber \\
  &  + \delta_{lm, l'm'}   \frac{1}{ 2\mu q^{2l+1}  C^2_l (\gamma) }    \frac{\Re \left ( G_{ l }^{(C,\infty)} (r, r' ; q) \right )  }{   (r r')^l }|_{r,r' \rightarrow 0}      . \label{Mzeta}
\end{align}  
Both $G_{ lm, l'm'}^{(C,t)}$ and $G_{ l }^{(C,\infty)}$ are   ultraviolet divergent, and  after cancellation between two terms,  generalized zeta function is finite and well-defined. At the limit of $\gamma \rightarrow 0$, 
\begin{equation}
C_l (\gamma) \stackrel{\gamma \rightarrow 0}{\rightarrow}  C_l (0)  =\frac{\sqrt{\pi}}{2^{l+1} \Gamma(l+ \frac{3}{2})} = \frac{j_l(qr)}{(q r)^l}|_{r\rightarrow 0},
\end{equation}
and
\begin{equation}
  \frac{\Re \left ( G_{ l }^{(C,\infty)} (r, r'' ; q) \right )  }{   (r r'')^l }|_{r,r'' \rightarrow 0}    \stackrel{\gamma \rightarrow 0}{\rightarrow}   2\mu q    \frac{ j_l(q r) n_l (q r)}{   r^{2l} }|_{r \rightarrow 0} ,
\end{equation}
hence,  Eq.(\ref{Mzeta}) is reduced to Eq.(B32) in Ref.~\cite{Guo:2021lhz}.

\section{Discussion and summary}\label{summary}

\subsection{Perturbation expansion}
The key element of  generalized zeta function in Eq.(\ref{Mzeta}) is Coulomb interaction modified Green's function in a trap, which is given by  Dyson equation, Eq.(\ref{DysonTCGF}). Solving Dyson equation in most cases is not an easy task, great effort must to be made to deal with both ultraviolet divergence (UV) and infrared divergence (IR) caused by Coulomb interaction. Therefore the perturbation expansion may be more practical in general,   see discussion in \cite{Beane:2014qha,Stellin:2020gst}. Symbolically, the Coulomb force modified zeta function is a real function and given by
 \begin{equation}
 \mathcal{\hat{M}}_{C,t} \sim \frac{1}{C^2(\gamma)} \left [ \hat{G}_{C,t} - \Re \left (\hat{ G}_{C,\infty} \right ) \right ],
 \end{equation}
 where the solution of  $\hat{G}_{C,t} $ is given by
 \begin{equation}
 \hat{G}_{C,t}  = \frac{\hat{G}_{t} }{1- \hat{V}_C \hat{G}_{t} } = \sum_{n=0}^\infty \hat{G}_{t} \left ( \hat{V}_C \hat{G}_{t} \right )^n,
 \end{equation}
 and  $ \hat{G}_{t} $ denotes Green's function in a trap,
  \begin{equation}
 \hat{G}_{t} (E)= \frac{1}{E- \hat{H}_t}.
 \end{equation}
Although the analytic expression of infinite volume Coulomb Green's function $\hat{ G}_{C,\infty}$ is known already, in order to make sure the UV and IR divergences cancelled out properly order by order, $\hat{ G}_{C,\infty}$ can be expanded in terms of perturbation theory as well
 \begin{equation}
 \hat{G}_{C,\infty}  = \frac{\hat{G}_{0,\infty} }{1- \hat{V}_C \hat{G}_{0,\infty} } = \sum_{n=0}^\infty \hat{G}_{0,\infty} \left ( \hat{V}_C \hat{G}_{0,\infty} \right )^n,
 \end{equation}
 where
 \begin{equation}
 \hat{G}_{0,\infty} (E)= \frac{1}{E- \hat{H}_0}.
 \end{equation}
 Hence, Coulomb corrected zeta function may be computed by perturbation expansion systematically, 
   \begin{align}
&C^2(\gamma)  \mathcal{\hat{M}}_{C}  \nonumber \\
& \sim  \sum_{n=0}^\infty \left [  \hat{G}_{t} \left ( \hat{V}_C \hat{G}_{t} \right )^n - \Re \left ( \hat{G}_{0,\infty} \left ( \hat{V}_C \hat{G}_{0,\infty} \right )^n \right ) \right ].
 \end{align}
Perturbation expansion may be well applied to both HO trap and periodic cubic box, also see \cite{Beane:2014qha,Stellin:2020gst} for the discussion in finite volume from effective field theory perspective.

\paragraph{HO trap: } for the harmonic oscillator trap, iterating Dyson equation in Eq.(\ref{DysonTCGFHO}) once, the leading order and first order perturbation result can be written down formally by,
   \begin{align}
  &  \frac{ C^2_l (\gamma)}{    C^2_l (0)}   \mathcal{M}^{(C,0th)}_{l }  (\varepsilon)  =  (-1)^{l+1} \left (  \frac{ 4\mu \omega}{q^2} \right )^{l+ \frac{1}{2}}         \frac{ \Gamma (\frac{l}{2} + \frac{3}{4} - \frac{ \varepsilon  }{2 \omega}) }{\Gamma ( \frac{1}{4}-\frac{l}{2} - \frac{\varepsilon }{2\omega} )}   ,
\end{align}
and
\begin{align}
& \frac{ C^2_l (\gamma)}{    C^2_l (0)}   \mathcal{M}^{(C,1st)}_{l }  (\varepsilon)  \nonumber \\
&  =-\frac{2^{2l+2}\Gamma^2(l+\frac{3}{2})}{ 2\mu q^{2l+1}  \pi }   \frac{\triangle G_l^{(C,1st)} (r, r' ; \varepsilon)  }{  ( r r' )^{l} } |_{r,r'\rightarrow 0} ,
\end{align}
 where
 \begin{align}
&\triangle G_l^{(C,1st)} (r, r' ; \varepsilon)   \nonumber \\
& = -   \int_{0}^\infty {r''}^2 d  r''  G_l^{(\omega)} (r, r'' ; \varepsilon)  \frac{Z}{r''}  G_l^{(\omega)} (r'', r' ; \varepsilon) \nonumber \\
&+ \Re  \int_{0}^\infty {r''}^2 d  r''  G_l^{(0,\infty)} (r, r'' ; q)  \frac{Z}{r''}  G_l^{(0,\infty)} (r'', r' ; q) .
\end{align}

\paragraph{Periodic cubic box: }  similarly, in finite volume, the leading order and first order perturbation result are given by 
 \begin{align}
  &\frac{C^2_l (\gamma) }{C^2_l (0) }  \mathcal{M}^{(C, 0th)}_{lm, l'm'}  (\varepsilon)  \nonumber \\
  & = -  \frac{1}{L^3} \sum_{ \mathbf{ p} \in \frac{2\pi \mathbf{ n}}{L}}^{ \mathbf{ n} \in \mathbb{Z}^3} \frac{p^{l+ l'}}{q^{2l+1}}      \frac{Y_{lm } (\mathbf{ \hat{p}})  Y^*_{l'm' } (\mathbf{ \hat{p}}) }{2\mu \varepsilon - \mathbf{ p}^2 }   \nonumber \\
  &  - \delta_{lm, l'm'}      \frac{2^{2l+1}  \Gamma(l+ \frac{1}{2}) \Gamma(l + \frac{3}{2})}{\pi}  \frac{1}{ (qr)^{2l+1}}  |_{r\rightarrow 0}    ,
\end{align}  
and
 \begin{align}
  &\frac{C^2_l (\gamma) }{C^2_l (0) }   \mathcal{M}^{(C,1st)}_{lm, l'm'}  (\varepsilon)   \nonumber \\
  &= \frac{1}{ 2\mu q^{2l+1}    }  \frac{1}{L^6} \sum_{ \mathbf{ p}, \mathbf{ p}' \in \frac{2\pi \mathbf{ n}}{L}}^{ \mathbf{ n} \in \mathbb{Z}^3}    \frac{p^l Y_{lm } (\mathbf{ \hat{p}}) }{\varepsilon - \frac{\mathbf{ p}^2}{2\mu}}  \frac{4\pi Z}{|\mathbf{ p} - \mathbf{ p}'|^2} \frac{{p'}^{l'} Y^*_{l'm' } (\mathbf{ \hat{p}}') }{\varepsilon - \frac{\mathbf{ p'}^2}{2\mu}}   \nonumber \\
  &  -   \frac{\delta_{lm, l'm'} }{ 2\mu q^{2l+1}   }  \Re \int \frac{d \mathbf{ p} d \mathbf{ p}'}{(2\pi)^6}   \frac{p^l Y_{lm } (\mathbf{ \hat{p}}) }{\varepsilon - \frac{\mathbf{ p}^2}{2\mu}}  \frac{4\pi Z}{|\mathbf{ p} - \mathbf{ p}'|^2} \frac{{p'}^{l} Y^*_{lm } (\mathbf{ \hat{p}}') }{\varepsilon - \frac{\mathbf{ p'}^2}{2\mu}}    . 
\end{align}

\subsection{Analytic solutions in a spherical hard wall trap}
The rotational symmetry inside of a hard-sphere trap is also well-preserved, so angular orbital momentum is still a good quantum number, only diagonal elements of Coulomb Green's function contribute. The partial wave Coulomb Green's function inside hard-sphere must be  the combination of regular and irregular functions of Coulomb differential equation: $ j_l^{(C)} ( \gamma, q r)  $ and $ n_l^{(C)} ( \gamma, q r)  $ defined in Eq.(\ref{jLC}) and Eq.(\ref{nLC}) respectively. Similar to the hard-sphere trap without Coulomb interaction, see Eq.(54) in Ref.\cite{Guo:2021uig}, the closed form of Coulomb force modified Green's function inside hard-sphere is given by
\begin{align}
G_{ l }^{(C,h.s.)} & (r, r'' ; \varepsilon)    = -  2 \mu  q j_l^{(C)} ( \gamma, q r_<)  j_l^{(C)} (\gamma, q r_>) \nonumber \\
& \times \left [  \frac{ n_l^{(C)} ( \gamma, q R) }{ j_l^{(C)} ( \gamma, q R) }  - \frac{n_l^{(C)} (\gamma, q r_>) }{j_l^{(C)} (\gamma, q r_>) } \right] .
\end{align}
The real part of Coulomb Green's function in infinite volume is given by
 \begin{equation}
\Re \left (G_{ l }^{(C,\infty)}  (r, r'' ; \varepsilon)   \right ) =   2 \mu  q j_l^{(C)} ( \gamma, q r_<)  n_l^{(C)} (\gamma, q r_>) .
\end{equation}
Hence, after UV cancellation in Eq.(\ref{Mzeta}), the analytic expression of Coulomb force modified generalized zeta function for hard-sphere trap is obtained
 \begin{equation}
    \mathcal{M}^{(C, h.s.)}_{lm, l'm'}  (\varepsilon)   =\delta_{lm, l'm'}   \frac{C^2_l (0)  }{     C^2_l (\gamma) }  \frac{ n_l^{(C)} ( \gamma, q R) }{ j_l^{(C)} ( \gamma, q R) }  .
\end{equation}
The   quantization condition in a hard-sphere  trap in presence of Coulomb interaction is given in a closed-form:
 \begin{equation}
      \cot \delta^{(S)}_l (q)   =  \frac{C^2_l (0) }{     C^2_l (\gamma) }  \frac{ n_l^{(C)} ( \gamma, q R) }{ j_l^{(C)} ( \gamma, q R) }  , \label{qchsw}
\end{equation} 
where   $ j_l^{(C)} ( \gamma, q r)  $ and $ n_l^{(C)} ( \gamma, q r)  $ are defined in Eq.(\ref{jLC}) and Eq.(\ref{nLC}) respectively.

\subsection{Summary}
In summary, we present a general  discussion on the topic of formulating quantization condition of trapped systems by  including long-range Coulomb interaction.    Although all the discussion are based on non-perturbative LS equation approach, in most cases,  the Coulomb force modified Green's function in a trap must be solved either numerically or by perturbation expansion.  In  special cases, such as  the spherical hard wall trap, the closed-form of quantization condition is obtained and given in Eq.(\ref{qchsw}).

\acknowledgments

We thank fruitful  discussion with Bingwei Long.  P.G. also acknowledges support from the Department of Physics and Engineering, California State University, Bakersfield, CA. The work  was   supported in part by the National Science Foundation  under Grant No. NSF PHY-1748958.

\appendix

\section{Formal scattering theory with short-range and Coulomb interactions}\label{scattcoulinf}

In this section,  the  formal scattering theory in presence of both a short-range interaction and a long-range Coulomb interaction is briefly discussed, the  complete discussion  can be found in Refs.~\cite{mott1985,GOL64}. The connection to trapped system is also briefly discussed symbolically.

\subsection{Coulomb force modified scattering amplitude  in infinite volume}
The infinite volume scattering amplitude in the presence of both Coulomb and short-range nuclear interactions is defined by
\begin{equation}
T_\infty = - \langle \Psi_0 |( \hat{V}_C + \hat{V}_S) | \Psi^{(+)} \rangle,
\end{equation}
where $| \Psi_0  \rangle$ stands for plane wave. The  $| \Psi^{(+)} \rangle$ is defined by LS equation,
\begin{equation}
    | \Psi^{(\pm)} \rangle = | \Psi_0  \rangle + \hat{G}_0 (E \pm i 0) (\hat{V}_C+\hat{V}_S) | \Psi^{(\pm)} \rangle,\label{LSCNeq}
\end{equation}
where
\begin{equation}
\hat{G}_0 (E \pm i 0) = \frac{1}{E - \hat{H}_0 \pm i 0}.
\end{equation}
Using Eq.(\ref{LSCNeq}) and also        LS equation for pure Coulomb interaction,
\begin{equation}
    \langle  \Psi_0 |  =    \langle  \Psi^{(-)}_C | -   \langle  \Psi^{(-)}_C | \hat{V}_C \hat{G}_0 (E + i 0)    ,
\end{equation}
the total infinite volume scattering amplitude $T_\infty$ can be rewritten as, also see Refs. \cite{mott1985,GOL64},
\begin{equation}
T_\infty = -  \langle \Psi^{(-)}_C |  \hat{V}_C | \Psi_0  \rangle   - \langle \Psi^{(-)}_C |  \hat{V}_S | \Psi^{(+)} \rangle  . \label{TCNinf}
\end{equation}

  The first term   in Eq.(\ref{TCNinf}) is   identified as pure Coulomb interaction scattering amplitude,
\begin{equation}
T_{C,\infty}=-  \langle \Psi^{(-)}_C |  \hat{V}_C | \Psi_0  \rangle= -  \langle\Psi_0   |  \hat{V}_C |   \Psi^{(+)}_C\rangle.  
\end{equation}
  The partial wave coulomb amplitude is parameterized by Coulomb phase shifts,  
  \begin{equation}
  T^{(C,\infty)}_l  \propto  \frac{e^{2 i \delta^{(C)}_{l}}-1}{2i}  ,
  \end{equation}
    where the Coulomb phase shift $\delta_l^{(C)}$ is defined in Eq.(\ref{phasecoulomb}).

 The second term in Eq.(\ref{TCNinf})  is  the result of short-range interaction in the presence of Coulomb interaction, using LS equation
\begin{equation}
| \Psi^{(+)} \rangle = | \Psi^{(+)}_C \rangle + \hat{G}_C (E+ i 0) V_S | \Psi^{(+)} \rangle ,
\end{equation}
 where
\begin{equation}
\hat{G}_C (E \pm i 0) = \frac{1}{E - \hat{H}_0 - \hat{V}_C \pm i 0},
\end{equation}
it can be shown rather straight-forwardly that second term satisfies  equation
\begin{align}
 &    -  \langle \Psi^{(-)}_C |  \hat{V}_S | \Psi^{(+)} \rangle  \nonumber \\
 &=  -  \langle \Psi^{(-)}_C |  \hat{V}_S | \Psi^{(+)}_C \rangle  -  \langle \Psi^{(-)}_C |  \hat{V}_S \hat{G}_C (E+ i 0) \hat{V}_S | \Psi^{(+)} \rangle . \label{VCSeq}
\end{align}
 Hence, it may be useful and more convenient  to define a Coulomb force modified scattering operator
\begin{equation}
    \hat{T}_{SC,\infty}  | \Psi_C^{(+)}  \rangle  = - \hat{V}_S | \Psi^{(+)} \rangle,
\end{equation}
thus, the second term in Eq.(\ref{TCNinf}) now can be written as
\begin{equation}
- \langle \Psi^{(-)}_C |  \hat{V}_S | \Psi^{(+)} \rangle  = \langle \Psi^{(-)}_C | \hat{T}_{SC,\infty} | \Psi^{(+)}_C \rangle .
\end{equation}
According to Eq.(\ref{VCSeq}), $\hat{T}_{SC,\infty}$ satisfies operator equation
\begin{equation}
    \hat{T}_{SC,\infty} = - \hat{V}_S  + \hat{V}_S  \hat{G}_C (E+ i 0)  \hat{T}_{SC,\infty} . \label{TSCinfappend}
\end{equation}

The total scattering amplitude is now given by
\begin{equation}
T_\infty = T_{C,\infty} +  \langle \Psi^{(-)}_C | \hat{T}_{SC,\infty} | \Psi^{(+)}_C \rangle  .  
\end{equation}
Given the fact that the symbolic solution of $\hat{T}_{SC,\infty}$ operator is given by
 \begin{equation}
 \hat{T}^{-1}_{SC,\infty}= -   \hat{V}_S^{-1} + \hat{G}_C(E+ i 0) ,  \label{TNLSeq}
 \end{equation}
   and   
 \begin{equation}
 \langle \Psi^{(-)}_C  | \Psi^{(+)}_C \rangle   = 1+2 i T_{C,\infty} = S_{C,\infty}  \propto e^{2 i \delta^{(C)}_l}
 \end{equation}
 is the pure Coulomb interaction $S$-matrix,    the partial wave expansion of $\langle \Psi^{(-)}_C | \hat{T}_{SC,\infty} | \Psi^{(+)}_C \rangle$ is conventionally parameterized by both Coulomb phase shift $\delta_l^{(C)}$ and  the short-range interaction  phase shift, $\delta^{(S)}_l$, see Refs.~\cite{mott1985,GOL64},  
 \begin{equation}
     \langle \Psi^{(-)}_C | \hat{T}_{SC,\infty} | \Psi^{(+)}_C   \rangle \propto e^{2 i \delta^{(C)}_l}   \frac{e^{2 i \delta^{(S)}_{l}}-1}{2i}.
 \end{equation}
Therefore, the partial wave total infinite volume scattering amplitude is thus defined by a total phase shift, 
\begin{equation}
\delta_l = \delta^{(S)}_{l}+\delta^{(C)}_{l},
\end{equation}
  and
\begin{equation}
T^{(\infty)}_l  \propto  \frac{e^{2 i  \delta_{l} }-1}{2i} = \frac{e^{2 i  \delta^{(C)}_{l} }-1}{2i}  + e^{2 i \delta^{(C)}_l}   \frac{e^{2 i \delta^{(S)}_{l}}-1}{2i} .
\end{equation}

\subsection{Charged particles in a trap in presence of Coulomb force}
In the trap, the Eq.(\ref{TSCinfappend}) is now modified to
\begin{equation}
    \hat{T}_{SC,t} = -\hat{V}_S +  \hat{V}_S  \hat{G}_{C,t}(E+ i 0) \hat{T}_{SC,t} \, ,
\end{equation}
where 
\begin{equation}
    \hat{G}_{C,t}(E \pm i 0) = \frac{1}{E - \hat{H}_t - \hat{V}_{C} \pm i 0}  ,
\end{equation}
is Coulomb Green's function in the trap,  and
$$ \hat{H}_t = \hat{H}_0 + \hat{V}_{t}$$  is trap Hamiltonian operator.
The quantization  condition including Coulomb interaction   thus is given by
\begin{equation}
\det \left [ V_S^{-1}  - G_{C,t}( E+ i 0 )   \right] =0. \label{coulqc}
\end{equation}

\subsection{Quantization condition including Coulomb interaction}

 Eliminating  $V^{-1}_S$ from  Eq.(\ref{TNLSeq})  and Eq.(\ref{coulqc}), the  quantization condition Eq.(\ref{coulqc})  thus can be rewritten as
 \begin{equation}
\det \left [ T_{SC,\infty}^{-1} +   G_{C,t }( E+ i 0 )   - G_C (E + i 0)  \right] =0.
\end{equation}
In general, Coulomb Green's function in the trap is obtained by Dyson equation,
\begin{equation}
   \hat{G}_{C,t}(E)  = \hat{G}_t (E) + \hat{G}_t (E) \hat{V}_{C}   \hat{G}_{C,t}(E) ,
\end{equation}
where
\begin{equation}
    \hat{G}_{t}(E \pm i 0) = \frac{1}{E - \hat{H}_t   \pm i 0}   .
\end{equation}
In practice, Coulomb effect may be treated as perturbation by summing over all the ladder diagrams generated by  Coulomb exchanges,
\begin{equation}
   \hat{G}_{C,t}(E)  = \hat{G}_t (E)  \sum_{n=0}^\infty \left (  \hat{V}_{C}  \hat{G}_t (E)   \right )^n .
\end{equation}

\bibliography{ALL-REF.bib,nuclph.bib}

\begin{thebibliography}{47}
\expandafter\ifx\csname natexlab\endcsname\relax\def\natexlab#1{#1}\fi
\expandafter\ifx\csname bibnamefont\endcsname\relax
  \def\bibnamefont#1{#1}\fi
\expandafter\ifx\csname bibfnamefont\endcsname\relax
  \def\bibfnamefont#1{#1}\fi
\expandafter\ifx\csname citenamefont\endcsname\relax
  \def\citenamefont#1{#1}\fi
\expandafter\ifx\csname url\endcsname\relax
  \def\url#1{\texttt{#1}}\fi
\expandafter\ifx\csname urlprefix\endcsname\relax\def\urlprefix{URL }\fi
\providecommand{\bibinfo}[2]{#2}
\providecommand{\eprint}[2][]{\url{#2}}

\bibitem[{\citenamefont{L{\"u}scher}(1991)}]{Luscher:1990ux}
\bibinfo{author}{\bibfnamefont{M.}~\bibnamefont{L{\"u}scher}},
  \bibinfo{journal}{Nucl. Phys.} \textbf{\bibinfo{volume}{B354}},
  \bibinfo{pages}{531} (\bibinfo{year}{1991}).

\bibitem[{\citenamefont{Busch et~al.}(1998)\citenamefont{Busch, Englert,
  Rza\.zewski, and Wilkens}}]{Busch98}
\bibinfo{author}{\bibfnamefont{T.}~\bibnamefont{Busch}},
  \bibinfo{author}{\bibfnamefont{B.-G.} \bibnamefont{Englert}},
  \bibinfo{author}{\bibfnamefont{K.}~\bibnamefont{Rza\.zewski}},
  \bibnamefont{and} \bibinfo{author}{\bibfnamefont{M.}~\bibnamefont{Wilkens}},
  \bibinfo{journal}{Found. Phys.} \textbf{\bibinfo{volume}{28}},
  \bibinfo{pages}{549–559} (\bibinfo{year}{1998}).

\bibitem[{\citenamefont{Rummukainen and Gottlieb}(1995)}]{Rummukainen:1995vs}
\bibinfo{author}{\bibfnamefont{K.}~\bibnamefont{Rummukainen}} \bibnamefont{and}
  \bibinfo{author}{\bibfnamefont{S.~A.} \bibnamefont{Gottlieb}},
  \bibinfo{journal}{Nucl. Phys.} \textbf{\bibinfo{volume}{B450}},
  \bibinfo{pages}{397} (\bibinfo{year}{1995}), \eprint{hep-lat/9503028}.

\bibitem[{\citenamefont{Christ et~al.}(2005)\citenamefont{Christ, Kim, and
  Yamazaki}}]{Christ:2005gi}
\bibinfo{author}{\bibfnamefont{N.~H.} \bibnamefont{Christ}},
  \bibinfo{author}{\bibfnamefont{C.}~\bibnamefont{Kim}}, \bibnamefont{and}
  \bibinfo{author}{\bibfnamefont{T.}~\bibnamefont{Yamazaki}},
  \bibinfo{journal}{Phys. Rev.} \textbf{\bibinfo{volume}{D72}},
  \bibinfo{pages}{114506} (\bibinfo{year}{2005}), \eprint{hep-lat/0507009}.

\bibitem[{\citenamefont{Bernard et~al.}(2008)\citenamefont{Bernard, Lage,
  Mei{\ss}ner, and Rusetsky}}]{Bernard:2008ax}
\bibinfo{author}{\bibfnamefont{V.}~\bibnamefont{Bernard}},
  \bibinfo{author}{\bibfnamefont{M.}~\bibnamefont{Lage}},
  \bibinfo{author}{\bibfnamefont{U.-G.} \bibnamefont{Mei{\ss}ner}},
  \bibnamefont{and} \bibinfo{author}{\bibfnamefont{A.}~\bibnamefont{Rusetsky}},
  \bibinfo{journal}{JHEP} \textbf{\bibinfo{volume}{08}}, \bibinfo{pages}{024}
  (\bibinfo{year}{2008}), \eprint{0806.4495}.

\bibitem[{\citenamefont{He et~al.}(2005)\citenamefont{He, Feng, and
  Liu}}]{He:2005ey}
\bibinfo{author}{\bibfnamefont{S.}~\bibnamefont{He}},
  \bibinfo{author}{\bibfnamefont{X.}~\bibnamefont{Feng}}, \bibnamefont{and}
  \bibinfo{author}{\bibfnamefont{C.}~\bibnamefont{Liu}},
  \bibinfo{journal}{JHEP} \textbf{\bibinfo{volume}{07}}, \bibinfo{pages}{011}
  (\bibinfo{year}{2005}), \eprint{hep-lat/0504019}.

\bibitem[{\citenamefont{Lage et~al.}(2009)\citenamefont{Lage, Mei{\ss}ner, and
  Rusetsky}}]{Lage:2009zv}
\bibinfo{author}{\bibfnamefont{M.}~\bibnamefont{Lage}},
  \bibinfo{author}{\bibfnamefont{U.-G.} \bibnamefont{Mei{\ss}ner}},
  \bibnamefont{and} \bibinfo{author}{\bibfnamefont{A.}~\bibnamefont{Rusetsky}},
  \bibinfo{journal}{Phys. Lett.} \textbf{\bibinfo{volume}{B681}},
  \bibinfo{pages}{439} (\bibinfo{year}{2009}), \eprint{0905.0069}.

\bibitem[{\citenamefont{D{\"o}ring et~al.}(2011)\citenamefont{D{\"o}ring,
  Mei{\ss}ner, Oset, and Rusetsky}}]{Doring:2011vk}
\bibinfo{author}{\bibfnamefont{M.}~\bibnamefont{D{\"o}ring}},
  \bibinfo{author}{\bibfnamefont{U.-G.} \bibnamefont{Mei{\ss}ner}},
  \bibinfo{author}{\bibfnamefont{E.}~\bibnamefont{Oset}}, \bibnamefont{and}
  \bibinfo{author}{\bibfnamefont{A.}~\bibnamefont{Rusetsky}},
  \bibinfo{journal}{Eur. Phys. J.} \textbf{\bibinfo{volume}{A47}},
  \bibinfo{pages}{139} (\bibinfo{year}{2011}), \eprint{1107.3988}.

\bibitem[{\citenamefont{Guo et~al.}(2013)\citenamefont{Guo, Dudek, Edwards, and
  Szczepaniak}}]{Guo:2012hv}
\bibinfo{author}{\bibfnamefont{P.}~\bibnamefont{Guo}},
  \bibinfo{author}{\bibfnamefont{J.}~\bibnamefont{Dudek}},
  \bibinfo{author}{\bibfnamefont{R.}~\bibnamefont{Edwards}}, \bibnamefont{and}
  \bibinfo{author}{\bibfnamefont{A.~P.} \bibnamefont{Szczepaniak}},
  \bibinfo{journal}{Phys. Rev.} \textbf{\bibinfo{volume}{D88}},
  \bibinfo{pages}{014501} (\bibinfo{year}{2013}), \eprint{1211.0929}.

\bibitem[{\citenamefont{Guo}(2013)}]{Guo:2013vsa}
\bibinfo{author}{\bibfnamefont{P.}~\bibnamefont{Guo}}, \bibinfo{journal}{Phys.
  Rev.} \textbf{\bibinfo{volume}{D88}}, \bibinfo{pages}{014507}
  (\bibinfo{year}{2013}), \eprint{1304.7812}.

\bibitem[{\citenamefont{Kreuzer and Hammer}(2009)}]{Kreuzer:2008bi}
\bibinfo{author}{\bibfnamefont{S.}~\bibnamefont{Kreuzer}} \bibnamefont{and}
  \bibinfo{author}{\bibfnamefont{H.~W.} \bibnamefont{Hammer}},
  \bibinfo{journal}{Phys. Lett.} \textbf{\bibinfo{volume}{B673}},
  \bibinfo{pages}{260} (\bibinfo{year}{2009}), \eprint{0811.0159}.

\bibitem[{\citenamefont{Polejaeva and Rusetsky}(2012)}]{Polejaeva:2012ut}
\bibinfo{author}{\bibfnamefont{K.}~\bibnamefont{Polejaeva}} \bibnamefont{and}
  \bibinfo{author}{\bibfnamefont{A.}~\bibnamefont{Rusetsky}},
  \bibinfo{journal}{Eur. Phys. J.} \textbf{\bibinfo{volume}{A48}},
  \bibinfo{pages}{67} (\bibinfo{year}{2012}), \eprint{1203.1241}.

\bibitem[{\citenamefont{Hansen and Sharpe}(2014)}]{Hansen:2014eka}
\bibinfo{author}{\bibfnamefont{M.~T.} \bibnamefont{Hansen}} \bibnamefont{and}
  \bibinfo{author}{\bibfnamefont{S.~R.} \bibnamefont{Sharpe}},
  \bibinfo{journal}{Phys. Rev.} \textbf{\bibinfo{volume}{D90}},
  \bibinfo{pages}{116003} (\bibinfo{year}{2014}), \eprint{1408.5933}.

\bibitem[{\citenamefont{Mai and D{\"o}ring}(2017)}]{Mai:2017bge}
\bibinfo{author}{\bibfnamefont{M.}~\bibnamefont{Mai}} \bibnamefont{and}
  \bibinfo{author}{\bibfnamefont{M.}~\bibnamefont{D{\"o}ring}},
  \bibinfo{journal}{Eur. Phys. J.} \textbf{\bibinfo{volume}{A53}},
  \bibinfo{pages}{240} (\bibinfo{year}{2017}), \eprint{1709.08222}.

\bibitem[{\citenamefont{Mai and Döring}(2019)}]{Mai:2018djl}
\bibinfo{author}{\bibfnamefont{M.}~\bibnamefont{Mai}} \bibnamefont{and}
  \bibinfo{author}{\bibfnamefont{M.}~\bibnamefont{Döring}},
  \bibinfo{journal}{Phys. Rev. Lett.} \textbf{\bibinfo{volume}{122}},
  \bibinfo{pages}{062503} (\bibinfo{year}{2019}), \eprint{1807.04746}.

\bibitem[{\citenamefont{D\"oring et~al.}(2018)\citenamefont{D\"oring, Hammer,
  Mai, Pang, Rusetsky, and Wu}}]{Doring:2018xxx}
\bibinfo{author}{\bibfnamefont{M.}~\bibnamefont{D\"oring}},
  \bibinfo{author}{\bibfnamefont{H.-W.} \bibnamefont{Hammer}},
  \bibinfo{author}{\bibfnamefont{M.}~\bibnamefont{Mai}},
  \bibinfo{author}{\bibfnamefont{J.-Y.} \bibnamefont{Pang}},
  \bibinfo{author}{\bibfnamefont{A.}~\bibnamefont{Rusetsky}}, \bibnamefont{and}
  \bibinfo{author}{\bibfnamefont{J.}~\bibnamefont{Wu}}, \bibinfo{journal}{Phys.
  Rev. D} \textbf{\bibinfo{volume}{97}}, \bibinfo{pages}{114508}
  (\bibinfo{year}{2018}), \eprint{1802.03362}.

\bibitem[{\citenamefont{Guo}(2017)}]{Guo:2016fgl}
\bibinfo{author}{\bibfnamefont{P.}~\bibnamefont{Guo}}, \bibinfo{journal}{Phys.
  Rev.} \textbf{\bibinfo{volume}{D95}}, \bibinfo{pages}{054508}
  (\bibinfo{year}{2017}), \eprint{1607.03184}.

\bibitem[{\citenamefont{Guo and Gasparian}(2017)}]{Guo:2017ism}
\bibinfo{author}{\bibfnamefont{P.}~\bibnamefont{Guo}} \bibnamefont{and}
  \bibinfo{author}{\bibfnamefont{V.}~\bibnamefont{Gasparian}},
  \bibinfo{journal}{Phys. Lett.} \textbf{\bibinfo{volume}{B774}},
  \bibinfo{pages}{441} (\bibinfo{year}{2017}), \eprint{1701.00438}.

\bibitem[{\citenamefont{Guo and Gasparian}(2018)}]{Guo:2017crd}
\bibinfo{author}{\bibfnamefont{P.}~\bibnamefont{Guo}} \bibnamefont{and}
  \bibinfo{author}{\bibfnamefont{V.}~\bibnamefont{Gasparian}},
  \bibinfo{journal}{Phys. Rev.} \textbf{\bibinfo{volume}{D97}},
  \bibinfo{pages}{014504} (\bibinfo{year}{2018}), \eprint{1709.08255}.

\bibitem[{\citenamefont{Guo and Morris}(2019)}]{Guo:2018xbv}
\bibinfo{author}{\bibfnamefont{P.}~\bibnamefont{Guo}} \bibnamefont{and}
  \bibinfo{author}{\bibfnamefont{T.}~\bibnamefont{Morris}},
  \bibinfo{journal}{Phys. Rev.} \textbf{\bibinfo{volume}{D99}},
  \bibinfo{pages}{014501} (\bibinfo{year}{2019}), \eprint{1808.07397}.

\bibitem[{\citenamefont{Mai et~al.}(2019)\citenamefont{Mai, Döring, Culver,
  and Alexandru}}]{Mai:2019fba}
\bibinfo{author}{\bibfnamefont{M.}~\bibnamefont{Mai}},
  \bibinfo{author}{\bibfnamefont{M.}~\bibnamefont{Döring}},
  \bibinfo{author}{\bibfnamefont{C.}~\bibnamefont{Culver}}, \bibnamefont{and}
  \bibinfo{author}{\bibfnamefont{A.}~\bibnamefont{Alexandru}}
  (\bibinfo{year}{2019}), \eprint{1909.05749}.

\bibitem[{\citenamefont{Guo et~al.}(2018)\citenamefont{Guo, Döring, and
  Szczepaniak}}]{Guo:2018ibd}
\bibinfo{author}{\bibfnamefont{P.}~\bibnamefont{Guo}},
  \bibinfo{author}{\bibfnamefont{M.}~\bibnamefont{Döring}}, \bibnamefont{and}
  \bibinfo{author}{\bibfnamefont{A.~P.} \bibnamefont{Szczepaniak}},
  \bibinfo{journal}{Phys. Rev.} \textbf{\bibinfo{volume}{D98}},
  \bibinfo{pages}{094502} (\bibinfo{year}{2018}), \eprint{1810.01261}.

\bibitem[{\citenamefont{Guo}(2020{\natexlab{a}})}]{Guo:2019hih}
\bibinfo{author}{\bibfnamefont{P.}~\bibnamefont{Guo}}, \bibinfo{journal}{Phys.
  Lett. B} \textbf{\bibinfo{volume}{804}}, \bibinfo{pages}{135370}
  (\bibinfo{year}{2020}{\natexlab{a}}), \eprint{1908.08081}.

\bibitem[{\citenamefont{Guo and D\"oring}(2020)}]{Guo:2019ogp}
\bibinfo{author}{\bibfnamefont{P.}~\bibnamefont{Guo}} \bibnamefont{and}
  \bibinfo{author}{\bibfnamefont{M.}~\bibnamefont{D\"oring}},
  \bibinfo{journal}{Phys. Rev. D} \textbf{\bibinfo{volume}{101}},
  \bibinfo{pages}{034501} (\bibinfo{year}{2020}), \eprint{1910.08624}.

\bibitem[{\citenamefont{Guo}(2020{\natexlab{b}})}]{Guo:2020wbl}
\bibinfo{author}{\bibfnamefont{P.}~\bibnamefont{Guo}}, \bibinfo{journal}{Phys.
  Rev.} \textbf{\bibinfo{volume}{D101}}, \bibinfo{pages}{054512}
  (\bibinfo{year}{2020}{\natexlab{b}}), \eprint{2002.04111}.

\bibitem[{\citenamefont{Guo and Long}(2020{\natexlab{a}})}]{Guo:2020kph}
\bibinfo{author}{\bibfnamefont{P.}~\bibnamefont{Guo}} \bibnamefont{and}
  \bibinfo{author}{\bibfnamefont{B.}~\bibnamefont{Long}},
  \bibinfo{journal}{Phys. Rev. D} \textbf{\bibinfo{volume}{101}},
  \bibinfo{pages}{094510} (\bibinfo{year}{2020}{\natexlab{a}}),
  \eprint{2002.09266}.

\bibitem[{\citenamefont{Guo}(2020{\natexlab{c}})}]{Guo:2020iep}
\bibinfo{author}{\bibfnamefont{P.}~\bibnamefont{Guo}}
  (\bibinfo{year}{2020}{\natexlab{c}}), \eprint{2007.04473}.

\bibitem[{\citenamefont{Guo and Long}(2020{\natexlab{b}})}]{Guo:2020ikh}
\bibinfo{author}{\bibfnamefont{P.}~\bibnamefont{Guo}} \bibnamefont{and}
  \bibinfo{author}{\bibfnamefont{B.}~\bibnamefont{Long}},
  \bibinfo{journal}{Phys. Rev. D} \textbf{\bibinfo{volume}{102}},
  \bibinfo{pages}{074508} (\bibinfo{year}{2020}{\natexlab{b}}),
  \eprint{2007.10895}.

\bibitem[{\citenamefont{Guo}(2020{\natexlab{d}})}]{Guo:2020spn}
\bibinfo{author}{\bibfnamefont{P.}~\bibnamefont{Guo}}, \bibinfo{journal}{Phys.
  Rev. D} \textbf{\bibinfo{volume}{102}}, \bibinfo{pages}{054514}
  (\bibinfo{year}{2020}{\natexlab{d}}), \eprint{2007.12790}.

\bibitem[{\citenamefont{Guo and Gasparian}(2021)}]{Guo:2021lhz}
\bibinfo{author}{\bibfnamefont{P.}~\bibnamefont{Guo}} \bibnamefont{and}
  \bibinfo{author}{\bibfnamefont{V.}~\bibnamefont{Gasparian}}
  (\bibinfo{year}{2021}), \eprint{2101.01150}.

\bibitem[{\citenamefont{Guo and Long}(2021)}]{Guo:2021uig}
\bibinfo{author}{\bibfnamefont{P.}~\bibnamefont{Guo}} \bibnamefont{and}
  \bibinfo{author}{\bibfnamefont{B.}~\bibnamefont{Long}}
  (\bibinfo{year}{2021}), \eprint{2101.03901}.

\bibitem[{\citenamefont{Kong and Ravndal}(2000)}]{Kong:1999sf}
\bibinfo{author}{\bibfnamefont{X.}~\bibnamefont{Kong}} \bibnamefont{and}
  \bibinfo{author}{\bibfnamefont{F.}~\bibnamefont{Ravndal}},
  \bibinfo{journal}{Nucl. Phys. A} \textbf{\bibinfo{volume}{665}},
  \bibinfo{pages}{137} (\bibinfo{year}{2000}), \eprint{hep-ph/9903523}.

\bibitem[{\citenamefont{Beane et~al.}(2020)}]{Beane:2020ycc}
\bibinfo{author}{\bibfnamefont{S.~R.} \bibnamefont{Beane}} \bibnamefont{et~al.}
  (\bibinfo{year}{2020}), \eprint{2003.12130}.

\bibitem[{\citenamefont{Beane and Savage}(2014)}]{Beane:2014qha}
\bibinfo{author}{\bibfnamefont{S.~R.} \bibnamefont{Beane}} \bibnamefont{and}
  \bibinfo{author}{\bibfnamefont{M.~J.} \bibnamefont{Savage}},
  \bibinfo{journal}{Phys. Rev. D} \textbf{\bibinfo{volume}{90}},
  \bibinfo{pages}{074511} (\bibinfo{year}{2014}), \eprint{1407.4846}.

\bibitem[{\citenamefont{Stellin and Mei\ss{}ner}(2021)}]{Stellin:2020gst}
\bibinfo{author}{\bibfnamefont{G.}~\bibnamefont{Stellin}} \bibnamefont{and}
  \bibinfo{author}{\bibfnamefont{U.-G.} \bibnamefont{Mei\ss{}ner}},
  \bibinfo{journal}{Eur. Phys. J. A} \textbf{\bibinfo{volume}{57}},
  \bibinfo{pages}{26} (\bibinfo{year}{2021}), \eprint{2008.06553}.

\bibitem[{\citenamefont{Messiah}(1999)}]{messiah1999quantum}
\bibinfo{author}{\bibfnamefont{A.}~\bibnamefont{Messiah}},
  \emph{\bibinfo{title}{Quantum Mechanics}}, Dover books on physics
  (\bibinfo{publisher}{Dover Publications}, \bibinfo{year}{1999}), ISBN
  \bibinfo{isbn}{9780486409245},
  \urlprefix\url{https://books.google.com/books?id=mwssSDXzkNcC}.

\bibitem[{\citenamefont{Rotureau et~al.}(2010)\citenamefont{Rotureau, Stetcu,
  Barrett, Birse, and van Kolck}}]{Rotureau:2010uz}
\bibinfo{author}{\bibfnamefont{J.}~\bibnamefont{Rotureau}},
  \bibinfo{author}{\bibfnamefont{I.}~\bibnamefont{Stetcu}},
  \bibinfo{author}{\bibfnamefont{B.}~\bibnamefont{Barrett}},
  \bibinfo{author}{\bibfnamefont{M.}~\bibnamefont{Birse}}, \bibnamefont{and}
  \bibinfo{author}{\bibfnamefont{U.}~\bibnamefont{van Kolck}},
  \bibinfo{journal}{Phys. Rev. A} \textbf{\bibinfo{volume}{82}},
  \bibinfo{pages}{032711} (\bibinfo{year}{2010}), \eprint{1006.3820}.

\bibitem[{\citenamefont{Rotureau et~al.}(2012)\citenamefont{Rotureau, Stetcu,
  Barrett, and van Kolck}}]{Rotureau:2011vf}
\bibinfo{author}{\bibfnamefont{J.}~\bibnamefont{Rotureau}},
  \bibinfo{author}{\bibfnamefont{I.}~\bibnamefont{Stetcu}},
  \bibinfo{author}{\bibfnamefont{B.}~\bibnamefont{Barrett}}, \bibnamefont{and}
  \bibinfo{author}{\bibfnamefont{U.}~\bibnamefont{van Kolck}},
  \bibinfo{journal}{Phys. Rev. C} \textbf{\bibinfo{volume}{85}},
  \bibinfo{pages}{034003} (\bibinfo{year}{2012}), \eprint{1112.0267}.

\bibitem[{\citenamefont{Luu et~al.}(2010)\citenamefont{Luu, Savage, Schwenk,
  and Vary}}]{Luu:2010hw}
\bibinfo{author}{\bibfnamefont{T.}~\bibnamefont{Luu}},
  \bibinfo{author}{\bibfnamefont{M.~J.} \bibnamefont{Savage}},
  \bibinfo{author}{\bibfnamefont{A.}~\bibnamefont{Schwenk}}, \bibnamefont{and}
  \bibinfo{author}{\bibfnamefont{J.~P.} \bibnamefont{Vary}},
  \bibinfo{journal}{Phys. Rev. C} \textbf{\bibinfo{volume}{82}},
  \bibinfo{pages}{034003} (\bibinfo{year}{2010}), \eprint{1006.0427}.

\bibitem[{\citenamefont{Zhang et~al.}(2020)\citenamefont{Zhang, Stroberg,
  Navr\'atil, Gwak, Melendez, Furnstahl, and Holt}}]{Zhang:2020rhz}
\bibinfo{author}{\bibfnamefont{X.}~\bibnamefont{Zhang}},
  \bibinfo{author}{\bibfnamefont{S.}~\bibnamefont{Stroberg}},
  \bibinfo{author}{\bibfnamefont{P.}~\bibnamefont{Navr\'atil}},
  \bibinfo{author}{\bibfnamefont{C.}~\bibnamefont{Gwak}},
  \bibinfo{author}{\bibfnamefont{J.}~\bibnamefont{Melendez}},
  \bibinfo{author}{\bibfnamefont{R.}~\bibnamefont{Furnstahl}},
  \bibnamefont{and} \bibinfo{author}{\bibfnamefont{J.}~\bibnamefont{Holt}},
  \bibinfo{journal}{Phys. Rev. Lett.} \textbf{\bibinfo{volume}{125}},
  \bibinfo{pages}{112503} (\bibinfo{year}{2020}), \eprint{2004.13575}.

\bibitem[{\citenamefont{Elhatisari et~al.}(2016)\citenamefont{Elhatisari, Lee,
  Mei\ss{}ner, and Rupak}}]{Elhatisari:2016hby}
\bibinfo{author}{\bibfnamefont{S.}~\bibnamefont{Elhatisari}},
  \bibinfo{author}{\bibfnamefont{D.}~\bibnamefont{Lee}},
  \bibinfo{author}{\bibfnamefont{U.-G.} \bibnamefont{Mei\ss{}ner}},
  \bibnamefont{and} \bibinfo{author}{\bibfnamefont{G.}~\bibnamefont{Rupak}},
  \bibinfo{journal}{Eur. Phys. J. A} \textbf{\bibinfo{volume}{52}},
  \bibinfo{pages}{174} (\bibinfo{year}{2016}), \eprint{1603.02333}.

\bibitem[{\citenamefont{Rokash et~al.}(2015)\citenamefont{Rokash, Pine,
  Elhatisari, Lee, Epelbaum, and Krebs}}]{Rokash:2015hra}
\bibinfo{author}{\bibfnamefont{A.}~\bibnamefont{Rokash}},
  \bibinfo{author}{\bibfnamefont{M.}~\bibnamefont{Pine}},
  \bibinfo{author}{\bibfnamefont{S.}~\bibnamefont{Elhatisari}},
  \bibinfo{author}{\bibfnamefont{D.}~\bibnamefont{Lee}},
  \bibinfo{author}{\bibfnamefont{E.}~\bibnamefont{Epelbaum}}, \bibnamefont{and}
  \bibinfo{author}{\bibfnamefont{H.}~\bibnamefont{Krebs}},
  \bibinfo{journal}{Phys. Rev. C} \textbf{\bibinfo{volume}{92}},
  \bibinfo{pages}{054612} (\bibinfo{year}{2015}), \eprint{1505.02967}.

\bibitem[{\citenamefont{Blinder}(1984)}]{Blinder83}
\bibinfo{author}{\bibfnamefont{S.}~\bibnamefont{Blinder}}, \bibinfo{journal}{J.
  Math. Phys.} \textbf{\bibinfo{volume}{25}}, \bibinfo{pages}{905}
  (\bibinfo{year}{1984}).

\bibitem[{{\relax DLMF}()}]{NIST:DLMF}
{\relax DLMF}, \emph{\bibinfo{title}{{\it NIST Digital Library of Mathematical
  Functions}}}, \bibinfo{howpublished}{http://dlmf.nist.gov/, Release 1.1.0 of
  2020-12-15}, \bibinfo{note}{f.~W.~J. Olver, A.~B. {Olde Daalhuis}, D.~W.
  Lozier, B.~I. Schneider, R.~F. Boisvert, C.~W. Clark, B.~R. Miller, B.~V.
  Saunders, H.~S. Cohl, and M.~A. McClain, eds.},
  \urlprefix\url{http://dlmf.nist.gov/}.

\bibitem[{\citenamefont{Hostler}(1964)}]{doi:10.1063/1.1704153}
\bibinfo{author}{\bibfnamefont{L.}~\bibnamefont{Hostler}},
  \bibinfo{journal}{Journal of Mathematical Physics}
  \textbf{\bibinfo{volume}{5}}, \bibinfo{pages}{591} (\bibinfo{year}{1964}),
  \eprint{https://doi.org/10.1063/1.1704153},
  \urlprefix\url{https://doi.org/10.1063/1.1704153}.

\bibitem[{\citenamefont{Mott and Massey}(1985)}]{mott1985}
\bibinfo{author}{\bibnamefont{Mott}} \bibnamefont{and}
  \bibinfo{author}{\bibnamefont{Massey}}, \emph{\bibinfo{title}{The Theory of
  Atomic Collisions}} (\bibinfo{publisher}{Oxford University Press},
  \bibinfo{year}{1985}), \bibinfo{edition}{3rd} ed., ISBN
  \bibinfo{isbn}{0198512422}.

\bibitem[{\citenamefont{Goldberger and Watson}(1964)}]{GOL64}
\bibinfo{author}{\bibfnamefont{M.~L.} \bibnamefont{Goldberger}}
  \bibnamefont{and} \bibinfo{author}{\bibfnamefont{K.~M.}
  \bibnamefont{Watson}}, \emph{\bibinfo{title}{Collision Theory}}
  (\bibinfo{publisher}{Wiley}, \bibinfo{address}{New York},
  \bibinfo{year}{1964}), ISBN \bibinfo{isbn}{0471311103}.

\end{thebibliography}

\end{document}